\documentclass[12pt,english,floatfix,superscriptaddress,aps,prd,preprint]{revtex4}
\usepackage[utf8]{inputenc}
\usepackage{amsmath}
\usepackage{amssymb}
\usepackage{amsbsy}
\usepackage{amsfonts}
\usepackage{amsopn}
\usepackage{amstext}
\usepackage{graphicx}
\usepackage{amssymb}
\usepackage{amsfonts}
\usepackage{amsmath}
\usepackage{amsmath,amsthm,amsfonts,amssymb}
\usepackage[mathcal]{eucal}
\usepackage{mathrsfs}
\usepackage{graphicx}
\usepackage[english]{babel}
\usepackage{color}
\usepackage{esint}
\usepackage[dvips]{epsfig}
\usepackage[dvips]{graphicx}
\usepackage{float}
\usepackage{units}
\usepackage{textcomp}

\usepackage{hyperref}
\usepackage{slashed}

\newcommand{\ie}{\begin{equation}}
\newcommand{\fe}{\end{equation}}
\newcommand{\se}{\begin{eqnarray}}
\newcommand{\ff}{\end{eqnarray}}

\begin{document}

\title{Thermodynamic properties of an Aharonov-Bohm quantum ring}


\author{R. R. S Oliveira}
\email{rubensrso@fisica.ufc.br}
\affiliation{Universidade Federal do Cear\'a (UFC), Departamento de F\'isica,\\ Campus do Pici, Fortaleza - CE, C.P. 6030, 60455-760 - Brazil.}


\author{A. A. Araújo Filho}
\email{dilto@fisica.ufc.br}
\affiliation{Universidade Federal do Cear\'a (UFC), Departamento de F\'isica,\\ Campus do Pici, Fortaleza - CE, C.P. 6030, 60455-760 - Brazil.}


\author{F. C. E. Lima}
\email{cleiton.estevao@fisica.ufc.br}
\affiliation{Universidade Federal do Cear\'a (UFC), Departamento de F\'isica,\\ Campus do Pici, Fortaleza - CE, C.P. 6030, 60455-760 - Brazil.}


\author{R. V. Maluf}
\email{r.v.maluf@fisica.ufc.br}
\affiliation{Universidade Federal do Cear\'a (UFC), Departamento de F\'isica,\\ Campus do Pici, Fortaleza - CE, C.P. 6030, 60455-760 - Brazil.}


\author{C. A. S. Almeida}
\email{carlos@fisica.ufc.br}
\affiliation{Universidade Federal do Cear\'a (UFC), Departamento de F\'isica,\\ Campus do Pici, Fortaleza - CE, C.P. 6030, 60455-760 - Brazil.}

\date{\today}

\begin{abstract}

In this paper, we investigate the thermodynamic properties of an Aharonov-Bohm (AB) quantum ring in a heat bath for both relativistic and non-relativistic cases. For accomplishing this, we used the partition function which was obtained numerically using the Euler-Maclaurin formula. In particular, we determined the energy spectra as well as the behavior of the main thermodynamic functions of the canonical ensemble, namely, the Helmholtz free energy, the mean energy, the entropy and the heat capacity. The so-called Dulong-Petit law was verified only for the relativistic case. We noticed that in the low energy regime, the relativistic thermodynamic functions are reduced to the non-relativistic case as well.

\end{abstract}

\maketitle

\section{Introduction}

In recent years, the study of quantum rings (QRs) has received much attention due to its variety of technological applications in single-photon emitters, nanoflash memories \cite{Nowozin,Fomin}, photonic detectors \cite{Michler,Fomin} and qubits for spintronic quantum computing \cite{Fomin}. Moreover, quantum rings (QRs) is a fruitful subject for studying topological influence in condensed matter physics \cite{Fomin}. They are remarkable nanostructures with a non-simply connected topology which gives rise an intriguing energy structure \cite{Fuhrer} which differs from most others low-dimensional systems such as quantum dots, quantum wires and quantum wells. Besides QRs are divided into two categories: the one-dimensional (1D) (rings of constant radius) \cite{Cheung,Meijer,Frustaglia,Lorke,Kettemann,Viefers,Splettstoesser,Souma,Chaplik} and the two-dimensional rings (2D) (rings of variable radius) \cite{Tan,Bulaev,Duque,Bakke,
Nowak}.


In particular, a special case of 1D QRs has obtained notoriety in the literature, the so-called Aharonov-Bohm (AB) rings \cite{Citro,Belich,Nitta}. Currently, there are several number of works that analyze the dynamics of AB rings in both theoretical and experimental approaches. For instance, AB rings are studied in connection with the Aharonov-Casher (AC) effect \cite{Aharonov3,Citro,Belich,Nitta}, Lorentz symmetry violation \cite{Belich}, mesoscopic decoherence \cite{Hansen}, electromagnetic resonator \cite{Reulet} and Rashba spin-orbit interaction \cite{Nitta,Aeberhard,Shelykh}.

The study of physical properties of such materials, focusing on their thermal properties, is of great interest in the condensed matter physics, especially in solid state. Indeed, it is justified by practical needs and the fundamental science knowledge \cite{Balandin}. It is worth mentioning that in Refs. \cite{Balandin,Pop,Alofi,Che,Ruoff} were investigated the thermal properties of graphene, graphite, carbon nanotubes and nanostructured carbon materials. Nevertheless, the energy spectra of relativistic and non-relativistic cases as well as the thermodynamic properties for a Dirac fermion confined in a AB ring are lack up to date .

The present work has the purpose of determining the thermal properties of an AB ring for the relativistic and non-relativistic cases. For accomplishing this, we consider a set of non-interacting indistinguishable $N$-fermions confined in a AB ring. Besides we noticed that the energy spectrum is non-degenerate which allowed us to consider the strength field approach and use the numerical method based on the Euler-MacLaurin formula to calculate the canonical partition function. Moreover, in the low energy regime, the relativistic thermodynamic functions are reduced to the non-relativistic case.

This work is organized as follows: In Section \ref{sec2}, we introduce the model of AB ring described by the Dirac equation which shows the Dirac spinor and the energy spectra of relativistic and non-relativistic cases. In Section \ref{sec3}, we determine explicitly the thermodynamic properties of the AB ring, such as the Helmholtz free energy, the mean energy, the entropy and the heat capacity. We discuss about the results as well. Finally, in Section \ref{conclusion}, we present our conclusions and the final remarks. 

\section{An AB ring modeled by (1+1)-dimensional Dirac equation \label{sec2}}

In this section, we obtain the relativistic and non-relativistic energy spectra of an AB ring modeled by the Dirac equation in polar coordinates whose signature of the metric tensor is $(+ - -)$. Introducing initially the Dirac equation in the $(3+1)$-dimensional Minkowski spacetime written in an orthogonal system (in natural units $\hbar=c=1$) \cite{Greiner}
\ie \left\{i\gamma^{\mu}D_{\mu}+\frac{i}{2}\sum_{k=1}^{3}\gamma^k\left[D_k\ \mathsf{ln}{\left(\frac{h_1 h_2 h_3}{h_k}\right)}\right]-m\right\}\psi(t,{\bf r})=0, \ \ (\mu=0,1,2,3)
\label{AB},\fe
where $\gamma^\mu$ are gamma matrices, $D_\mu=\frac{1}{h_\mu}\partial_\mu$ are the derivative, $h_k$ are scale factors of the corresponding to coordinate system, $m$ is the rest mass and $\psi(t,{\bf r})$ is the Dirac spinor. In polar coordinates $(t,\rho,\theta)$, the scale factors are $h_0=1$, $h_1=1$, $h_2=\rho=a$ and $h_3=1$, where $a=$ const is the radius of the ring for modeling a 1D circular ring \cite{Fuhrer,Cheung,Meijer,Frustaglia,Lorke,Kettemann}. In this sense, Eq. \eqref{AB} turns out to be
\ie i\frac{\partial\psi(t,\theta)}{\partial t}=(\alpha^2 p_\theta+\gamma^0 m)\psi(t,\theta)
\label{AB1},\fe
where $p_\theta=-\frac{i}{a}\frac{\partial}{\partial\theta}$, and $\theta$ is the azimuthal angel which lies in $0\le\theta\le2\pi$.

Considering that the AB ring admits stationary states whose two-component Dirac spinor are given by $\psi(t,\theta)=e^{-iEt}f(\theta)$, and that the electromagnetic minimal coupling, Eq. \eqref{AB1} becomes 
\ie Ef(\theta)=H_{ring}f(\theta)=\left[\alpha^2\left(p_\theta-qA_\theta\right)+\gamma^0 m\right]f(\theta)
\label{AB2}\fe
where $E$ is the relativistic total energy of fermion with electric charge $q$.

Since we are working in a $(1+1)$-dimensional Minkowski spacetime, it is convenient to define the Dirac matrices $\alpha^2$ and $\beta$ in terms of the Pauli matrices, i.e., $\alpha^2=-\alpha_2=-\sigma_2$ and $\gamma^0=\sigma_3$. Writing $f(\theta)=(R^+(\theta), R^-(\theta))^T$, we transform Eq. \eqref{AB2} as follows
\ie \left(
    \begin{array}{cc}
      m & \ \ \frac{1}{a}\frac{d}{d\theta}-iqA_\theta  \\
      -\frac{1}{a}\frac{d}{d\theta}+iqA_\theta & \ -m \\
    \end{array}
  \right)\left(
           \begin{array}{c}
             R^+(\theta)  \\
             R^-(\theta)  \\
           \end{array}
         \right)=E
         \left(\begin{array}{c}
             R^+(\theta)  \\
             R^-(\theta)  \\
           \end{array}
             \right)
\label{AB3}.\fe

To configure an AB ring we must use the vector potential of the AB effect for a fermion restricted to move in a circle of radius $a$. Explicitly, this vector potential is written as ${\bf A}=\frac{\Phi}{2\pi a}\hat{e}_\theta$, where $\Phi=\pi b^2 B$ is the magnetic flux of the solenoid of radius $b$ and $B$ is the constant magnetic field inside the solenoid \cite{Aharonov}. We transform Eq. \eqref{AB3} in a system of two first-order coupled differential equations given by
\ie (E-m)R^+(\theta)=\frac{1}{a}\left(\frac{d}{d\theta}-\frac{\Phi}{\Phi_0}\right)R^-(\theta)
\label{AB5},\fe
\ie (E+m)R^-(\theta)=\frac{1}{a}\left(-\frac{d}{d\theta}+\frac{\Phi}{\Phi_0}\right)R^+(\theta)
\label{AB6},\fe
where $\Phi_0\equiv{\frac{2\pi}{q}}$ is the magnetic flux quantum. Now, Substituting (6) into (5), we obtain a linear differential equation with constant coefficients
\ie \left[\frac{d^2}{d\theta^2}-2i\frac{\Phi}{\Phi_0}\frac{d}{d\theta}+a^2(E^2-m^2)-\frac{\Phi^2}{\Phi^2_0}\right]R^+(\theta)=0
\label{AB7}\fe
with normalized solutions are given by $R^+(\theta)=\frac{1}{\sqrt{2\pi}}e^{i\lambda\theta}$. Consequently, the coefficient $\lambda$ is
\ie \lambda_{\pm}=\frac{\Phi}{\Phi_0}\pm a\sqrt{E^2-m^2}
\label{AB9}.\fe

However, since $R^+(\theta)$ satisfies the periodicity condition $R^+(\theta+2\pi)=R^+(\theta)$, it requires that $\lambda_{\pm}$ is an integer number. In this sense, we obtain the following relativistic energy spectrum and the spinor for the Dirac fermion confined in an AB ring:
\ie 
\begin{split}
&E_{n,s}=\pm\sqrt{m^2+\frac{1}{a^2}\left(n-s\frac{\Phi}{\Phi_0}\right)^2}, \ \ (n=0,\pm 1, \pm 2, \ldots)\\
& \psi(t,\theta) = \frac{e^{i(\lambda \theta -Et)}}{\sqrt{2 \pi}} \left(
    \begin{array}{cc}
      1 \\ 
        -i \nu\lambda + \nu\frac{\Phi}{\Phi_{0}}\\
    \end{array}
  \right)
\label{energyspectrum},
\end{split}
\fe
where $\nu = \frac{1}{a(E+m)}$ and $s=\pm 1$. This parameter comes from the following relation: $\Phi_0\to{s\Phi_0}$, where $\Phi_0=\frac{2\pi}{e}>0$, and $s=+1$ is regarded to be a positive charged fermion, $s=-1$ to a negative charged fermion and $e>0$ is the elementary electric charge.

Now, using the prescription $E=\epsilon+m$ in \eqref{energyspectrum}, where $\epsilon<<m$, we obtain the following non-relativistic energy spectrum of a Dirac fermion confined in an AB ring \cite{Griffiths}
\ie \epsilon_{n,s}=\frac{1}{2ma^2}\left(n-s\frac{\Phi}{\Phi_0}\right)^2, \ \ (n=0,\pm 1, \pm 2, \ldots)
\label{energyspectrum2}.\fe

We see that the relativistic and non-relativistic energy spectra are non-degenerated. However, in the limit $\Phi\to{0}$, the energy spectra have a twofold degeneracy, i.e., each energy level is doubly degenerate (except for $n=0$). Moreover, positive $n$ represents a fermion traveling in the same direction of the current of the solenoid and negative $n$ describes a fermion traveling in the opposite direction.

\section{Thermodynamic properties of the AB ring \label{sec3}} 

In this section, we calculate the thermodynamic properties of the AB ring in contact with a thermal reservoir at finite temperature for the relativistic and non-relativistic cases. These properties are given by following thermodynamic quantities, namely, the Helmholtz free energy, the mean energy, the entropy and the heat capacity. For the sake of simplicity we assume that only fermions with positive energy ($E>0$), negatively charged ($s=-1$) are regarded to constitute the thermodynamic ensemble. 

\subsection{The relativistic case\label{subsec1}}

For calculating the thermodynamic properties, let us begin defining initially the fundamental object in statistical mechanics, the so-called partition function $Z$. Given the non-degenerate energy spectrum, we can define it by a sum over all possible states of the system
\ie Z_{1}=\sum_{n=0}^{\infty}e^{-\beta E_n}
\label{partition1},\fe
where $\beta=\frac{1}{k_B T}$, and $k_B$ the Boltzmann constant and $T$ is the thermodynamic equilibrium temperature. After we obtain $Z_{1}$, all thermodynamic properties of the AB ring can be addressed. The main thermodynamic functions of our interest are the Helmholtz free energy $F$, the mean energy $U$, the entropy $S$ and the heat capacity $C_V$ which are defined as follows
\ie F=-\frac{1}{\beta}\ \mathsf{ln}\ Z_N, \,\,\,\,
U=-\frac{\partial}{\partial\beta}\ \mathsf{ln}\ Z_N, \,\,\,\, S=k_B\beta^2\frac{\partial F}{\partial\beta},\,\,\,\, C_V=-k_B\beta^2\frac{\partial U}{\partial\beta} \label{ii},\fe
where $Z_N$ is the total partition function for an set of non-interacting indistinguishable $N$-fermions.

In particular, let us consider the partition function regarding the relativistic case. Using the expressions \eqref{energyspectrum} and \eqref{partition1}, one obtains that for a one-fermion confined in the AB ring the partition function is
\ie Z_1=\sum_{n=0}^{\infty}e^{-\beta\sqrt{A n^2+Bn+C}}
\label{partition2},\fe
where $A=\frac{1}{a^2}$, $B=\frac{1}{a^2}\frac{\Phi}{\Phi_0}$ and $C=1+\left(\frac{\Phi}{a\Phi_0}\right)^2$.

Since Eq. (\ref{partition2}) cannot be evaluated in a closed form, we assume that AB ring is submitted to strong enough magnetic field ($\Phi\gg\Phi_0$) and the above equation turns out to be $f(n)\simeq e^{-\beta\sqrt{Bn+C}}$, which is a monotonically decreasing function and the associated integral
\ie I(\beta)\simeq\int_{0}^{\infty}e^{-\beta\sqrt{Bx+C}} dx=\frac{2}{B\beta^2}(1+\beta\sqrt{C})e^{-\beta\sqrt{C}}
\label{integral}\fe
is convergent and may be evaluated. With the sake of calculating it, now, let us invoke the so-called Euler-MacLaurin sum formula \cite{Greiner1995}
\ie Z_1=\sum_{n=0}^{\infty}f(n)\simeq\frac{1}{2}f(0)+\int_{0}^{\infty}f(x) dx-\sum_{p=1}^{\infty}\frac{1}{(2p)!}B_{2p}f^{(2p-1)}(0)
\label{partition3},\fe
which may be rewritten simply as
\ie Z_1=\sum_{n=0}^{\infty}f(n)\simeq\frac{1}{2}f(0)+\int_{0}^{\infty}f(x) dx-\frac{1}{12}f'(0)+\frac{1}{720}f'''(0)-\ldots+
\label{partition4},\fe
where $B_{2p}$ are the Bernoulli numbers. Explicitly, above equation takes the form
\ie Z_1\simeq e^{-\beta\sqrt{C}}\left[\frac{2}{B\beta^{2}}(1+\beta\sqrt{C})+\frac{1}{2}+\left(\frac{B}{24\sqrt{C}}-\frac{B^3}{720\sqrt{C^5}}\right)\beta+\frac{1}{90}\left(\frac{A}{2C}-\frac{B^2}{8C^2}\right)\beta^2-\mathcal{O}(\beta^{3})\right]
\label{partition5}. \fe
where $\mathcal{O}(\beta^{3})$ are terms involving higher order terms of $\beta$ which will be neglected henceforth. Notice that when one considers the high temperatures regime ($\beta\ll 1$), Eq. \eqref{partition5} becomes
\ie Z_1\simeq\frac{2}{B\beta^{2}}(1+\beta\sqrt{C})
\label{partition6},\fe
which straightforwardly yields
\ie Z_N\simeq\left[\frac{2}{B\beta^{2}}(1+\beta\sqrt{C})\right]^N
\label{partition7}.\fe

Therefore, using the partition function \eqref{partition7}, the Helmholtz free energy, the mean energy, the entropy and the heat capacity for the relativistic case are written as follows
\ie F\simeq -\frac{N}{\beta}\ \mathsf{ln}\left[\frac{2}{B\beta^{2}}(1+\beta\sqrt{C})\right], \ \ U\simeq N\frac{(2+\beta\sqrt{C})}{(\beta+\beta^2\sqrt{C})}
\label{properties},\fe
\ie S\simeq Nk_B\left\{\mathsf{ln}\left[\frac{2}{B\beta^{2}}(1+\beta\sqrt{C})\right]+\frac{(2+\beta\sqrt{C})}{(1+\beta\sqrt{C})}\right\}, \ \ C_V\simeq Nk_B\left[\frac{2+4\beta\sqrt{C}+\beta^2C}{(1+\beta\sqrt{C})^2}\right]
\label{properties2}.\fe

\subsection{The non-relativistic case\label{subsec3}}

Now, let us take into account the non-relativistic case where most of phenomena in condensed matter physics take place.  Substituting our previous result of the energy spectrum \eqref{energyspectrum2} in the partition function \eqref{partition1} and considering the condition where $\Phi\gg\Phi_0$, we obtain
\ie Z_1\simeq\sum_{n=0}^{\infty}e^{-\beta(\bar{B}n+\bar{C})}
\label{partition8}\fe
where $\bar{B}=\frac{\Phi}{ma^2\Phi_0}$ and $\bar{C}=\left(\frac{\Phi}{\sqrt{2m}a\Phi_0}\right)^2$. In an explicit form, above expression is given by
\ie Z_1\simeq\frac{e^{-\beta\bar{C}}}{1-e^{-\beta\bar{B}}}
\label{partition9} \fe
and the total partition function for a set of non-interacting indistinguishable $N$-fermions is \ie Z_N\simeq\left[\frac{e^{-\beta\bar{C}}}{1-e^{-\beta\bar{B}}}\right]^N
\label{partition10}.\fe

Therefore, taking into account \eqref{partition10} and the previous definition in (\ref{ii}) the required thermodynamic functions, namely, the Helmholtz free energy, the mean energy, the entropy and the heat capacity for the non-relativistic case are written as
\ie \bar{F}\simeq-\frac{N}{\beta}\ \mathsf{ln}\left(\frac{e^{-\beta\bar{C}}}{1-e^{-\beta\bar{C}}}\right), \ \  \bar{S}\simeq Nk_B\left[\beta\left(\bar{C}+\frac{\bar{B}}{e^{\beta\bar{B}}-1}\right)+\mathsf{ln}\left(\frac{e^{\beta(\bar{B}-\bar{C})}}{e^{\beta\bar{B}}-1}\right)\right]
\label{properties3},\fe
\ie \bar{U}\simeq N\left(\frac{\bar{B}}{e^{\beta\bar{B}}-1}+\bar{C}\right), \ \ \bar{C}_V\simeq Nk_B\beta^2\left[\frac{\bar{B}^2 e^{\beta\bar{B}}}{(1-e^{\beta\bar{B}})^2}\right]
\label{properties4}.\fe

In what follows, all discussions and remarks concerned to both the relativistic and the non-relativistic cases are presented in the next section. The construction and the analysis of graphics are provided as well to elucidate the behavior of the thermodynamic functions calculated previously. Finally, we make our final remarks in the conclusion. 

\subsection{Results and discussions\label{subsec3}}

At the beginning, we displays our results on the calculation of the thermodynamic functions (\ref{ii}) for the relativistic case. Since we obtained a non-degenerate spectrum, for obtaining such results, we used the Euler-MacLaurin formula regarding a strong field approach which was sufficed to evaluate the partition function numerically. Nevertheless, we also provide the calculations of the same thermodynamic functions pointed out in (\ref{ii}) considering the non-relativistic case. We make a brief comparison between them highlighting the main features.

Here, we plot all profiles of the thermal quantities vs. $\tau$ for different values of the magnetic flux $\Phi$, namely, $\Phi = 50\Phi_{0}, 100\Phi_{0}, 150\Phi_{0}, 200\Phi_{0}$ which is displayed in Figs. 1 and 2. Considering the relativistic case, we see from Fig. 1 that the Helmontz function $F(T)/N$ has a little increasement when $\tau$ starts to increase. Nevertheless, in general it decreases for high values of $\tau$ and has bigger values when the magnetic flux decreases. In the interval $0\leqslant \tau \leqslant 2 $ the behavior of the function $U(T)/N$ does not depend on the different magnetic fluxes while in the interval $2\leqslant \tau \leqslant 20$ it starts increasing with a nearly linear behavior. The entropy $S(T)/N$ is slowly increasing for large $\tau$ and decreases for large magnetic fluxes $\Phi 's$. The heat capacity $C_{V}(T)/N$ tends to an asymptotic behavior fixed in the value $2$ when $\tau$ increases.

In addition, for the non-relativistic case, we see from Fig. 2 that the Helmontz function $\bar{F}(T)/N$ in the interval $0 \leqslant \tau \leqslant 10000$ decreases with a nearly linear behavior when $\tau$ increases and has bigger values when the magnetic flux $\Phi$ increases. The mean energy $\bar{U}(T)/N$ increases with a entirely linear behavior when $\tau$ increases which is similar when one compares with the relativistic case. The entropy $\bar{S}(T)/N$ as in the relativistic case, is low increasing. The heat capacity $\bar{C}_{V}(T)/N$ differently to the relativistic case, has a abrupt asymptotic behavior in the value 1.

In general, analyzing the thermodynamic properties for the relativistic and non-relativistic cases we observed that the average energy in both systems presents a critical point for low temperature, that is, close to T = 0 K. However, in both cases since the temperature of the system increases, the amount of free energy available to perform the ensemble work decreases rapidly. In contrast, we noted that when this ensemble tends to the thermal equilibrium in the reservoir, the mean energy of the system increases continuously. The thermal variation of the AB ring tends to increase until reaching the thermal equilibrium in both cases, once the equilibrium is reached, this quantity remains constant. We also observed that the thermodynamic quantities studied have a higher intensity in the relativistic case, that is, all the amounts assume considerable values for low temperatures, as predicted in the literature. These results are due to the fact that in the relativistic case we have more energetic quantum system than in the non-relativistic case. Moreover, based on the solid state physics, we noted that the well-known Dulong-Petit law is satisfied for the relativistic case, namely $C_{V}(T\rightarrow 0)/N\simeq 2K_{B}$. On the other hand, for the non-relativistic case we obtained only $\bar{C}_{V}(T\rightarrow 0)/N\simeq K_{B}$ which does not satisfy the well-known Dulong-Petit law.

\begin{figure}[ht]
\centering
\includegraphics[width=8cm,height=5.5cm]{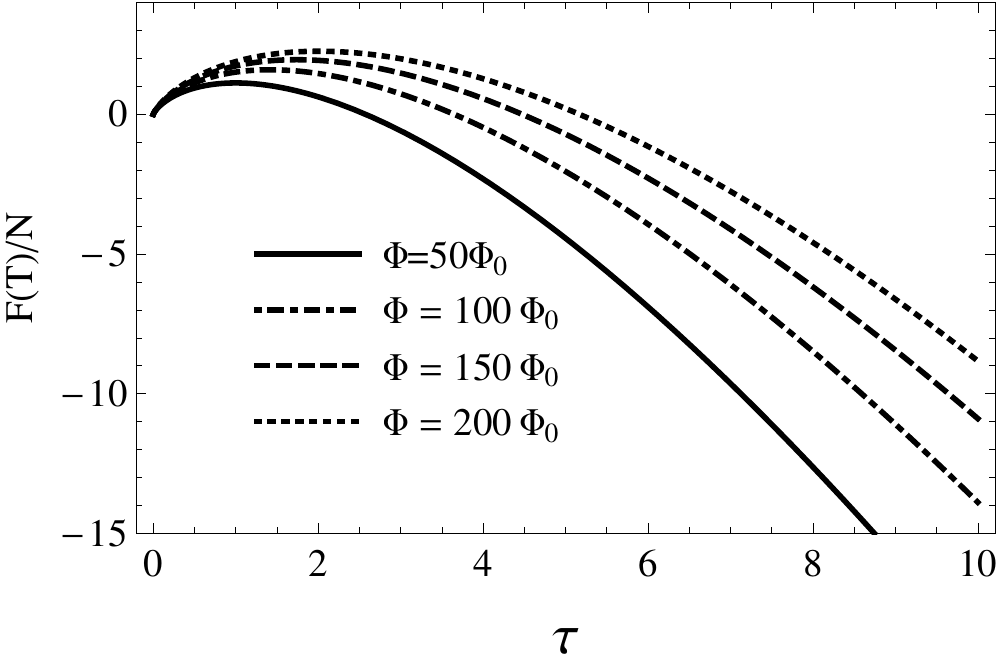}
\includegraphics[width=8cm,height=5.5cm]{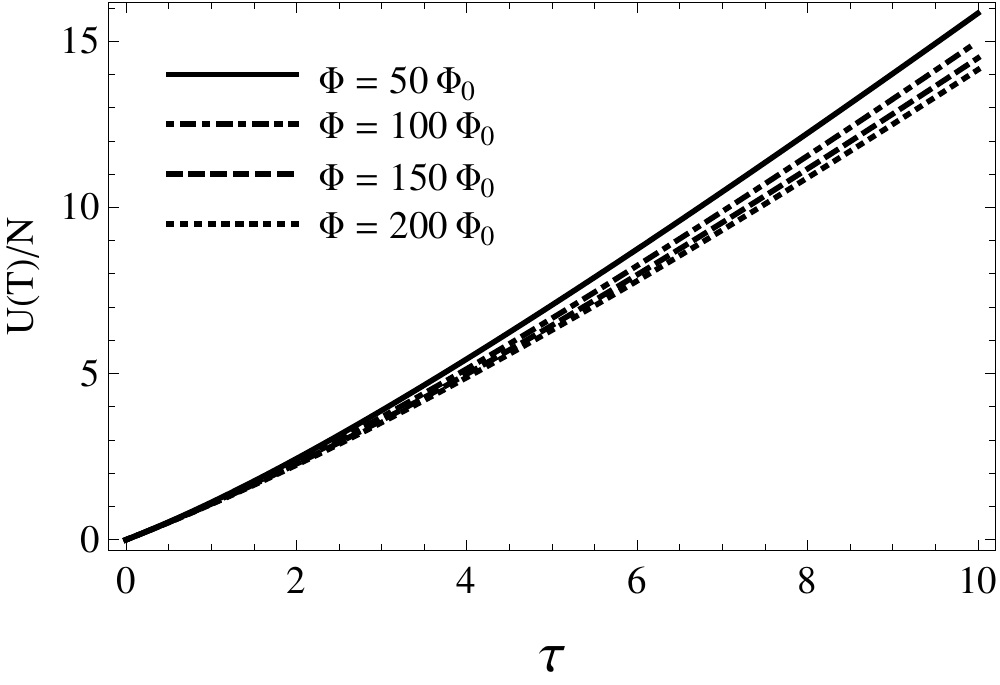}
\includegraphics[width=8cm,height=5.5cm]{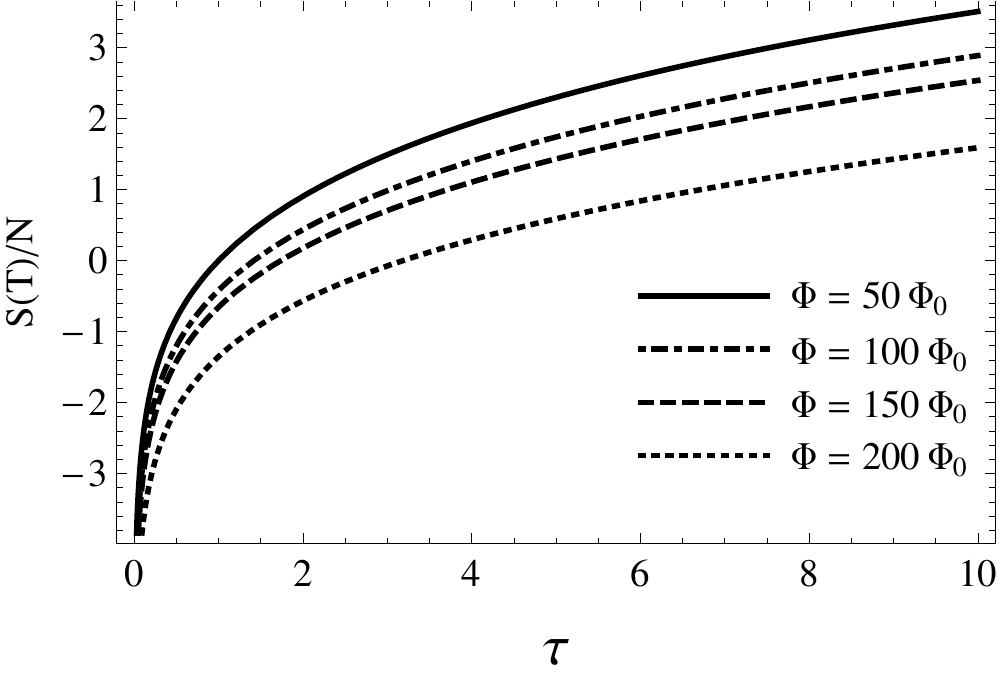}
\includegraphics[width=8cm,height=5.5cm]{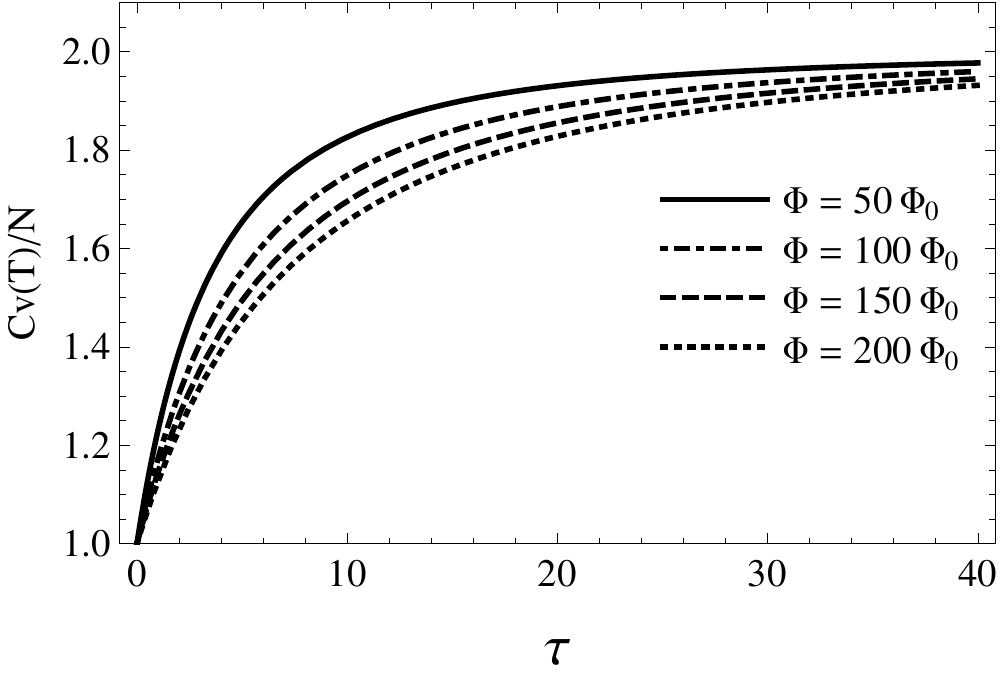}
\caption{Numerical solutions for the thermodynamic properties of the Aharanov-Bohm ring for the relativistic case: (a) The Helmholtz free energy; (b) the mean energy; (c) the entropy and (d) the heat capacity. In this case, we consider $\tau=K_{B}T$.}
\end{figure}


\begin{figure}[ht]
\centering
\includegraphics[width=8cm,height=5.5cm]{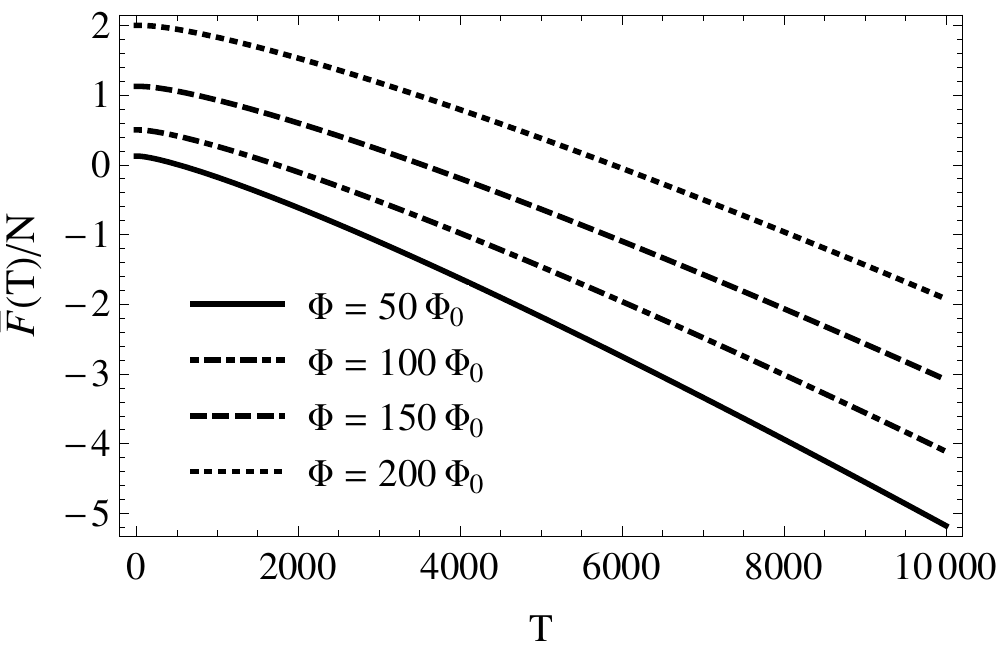}
\includegraphics[width=8cm,height=5.5cm]{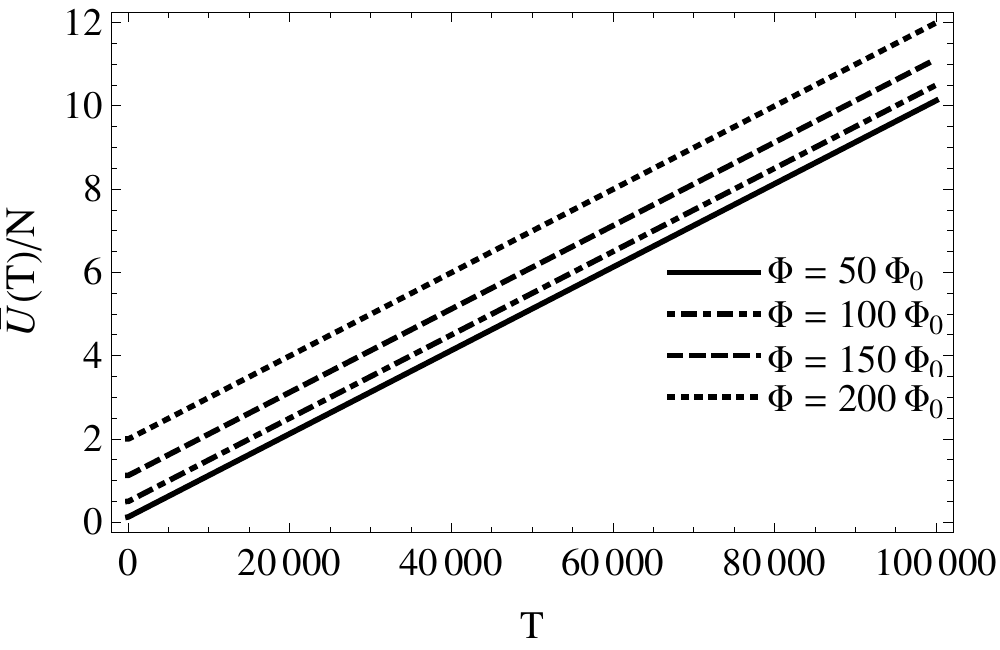}
\includegraphics[width=8cm,height=5.5cm]{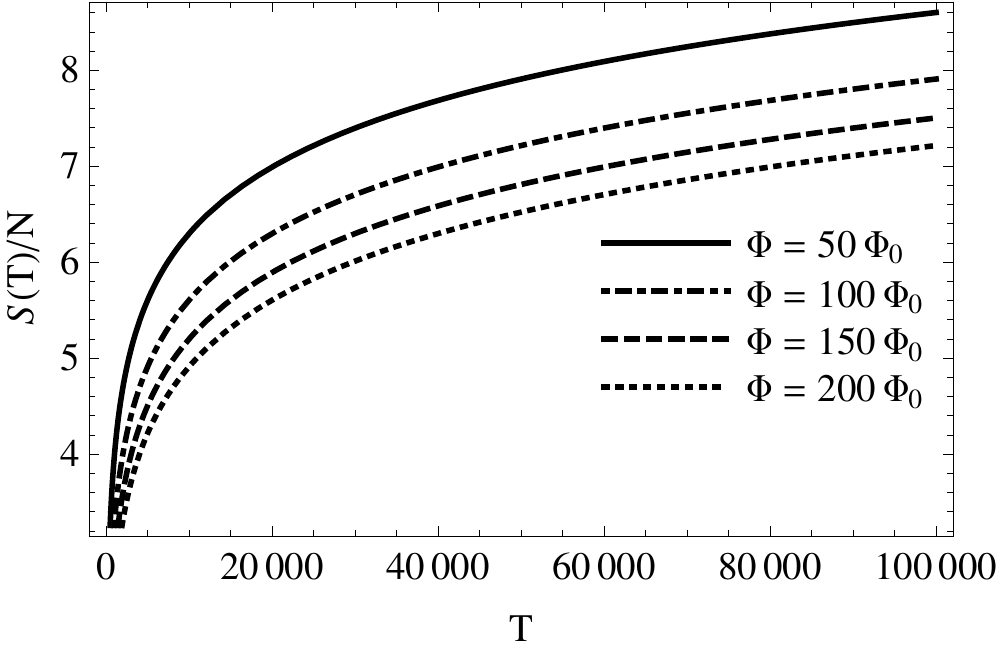}
\includegraphics[width=8cm,height=5.5cm]{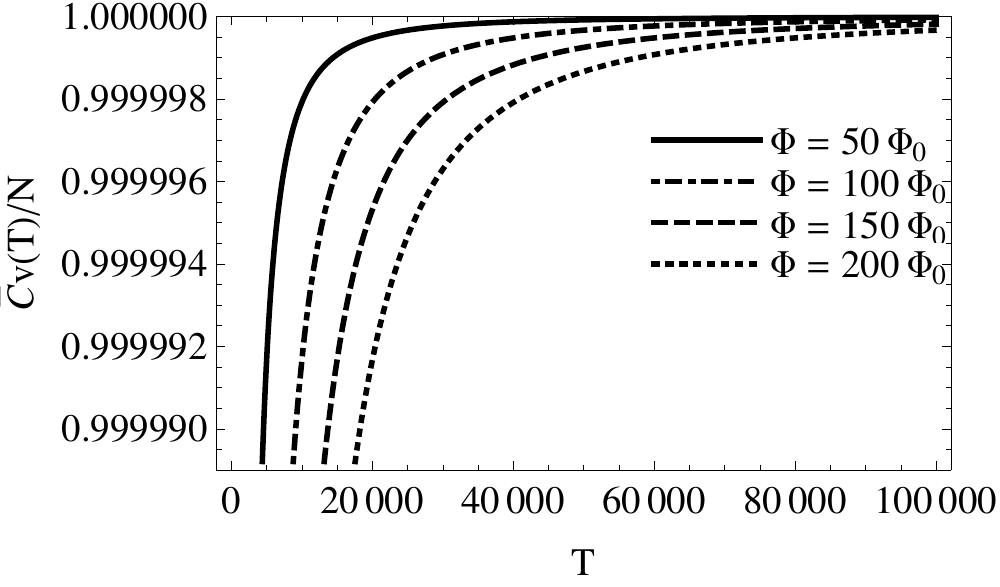}
\caption{Numerical solutions for the thermodynamic properties of the Aharanov-Bohm ring for the non-relativistic case: (a) the Helmholtz free energy; (b) the mean energy; (c) the entropy and (d) the heat capacity.}
\end{figure}

%
%

\section{Conclusion\label{conclusion}}

This work has the purpose of investigating the thermodynamic properties of an Aharovov-Bohm quantum ring in a heat bath regarding both relativistic and non-relativistic cases in the canonical ensemble framework at finite temperature. First, we calculated the relativistic energy spectrum and the spinor for relativistic case and we took a low energy limit where non-relativistic energy spectrum was recovered. It is worth mentioning that the energy spectrum turned out to be non-degenerated. Since the entire partition function did not have a closed form for carrying out our calculations, we performed it numerical regarding the strong field approximation which turned out to have a smooth behavior when the parameter $\tau$ started to increase. Moreover, when a low energy limit is taken into account, we obtain the non-relativistic regime as expected. In possession of the partition function, all the main thermodynamic properties could be derived, namely the Helmontz function $F(T)$, the mean energy $U(T)$, the entropy $S(T)$ and the heat capacity $C_{V}(T)$ and their respective graphics were plotted for several values of $\tau$ and different magnetic fluxes $\Phi \, 's$.

We noticed that the well-known Dulong-Petit law is satisfied for the relativistic case, namely $C_{V}(T\rightarrow 0)/N\simeq 2K_{B}$. On the other hand, for the non-relativistic case we obtained only $\bar{C}_{V}(T\rightarrow 0)/N\simeq K_{B}$ which does not satisfy the well-known Dulong-Petit law. In this way, we conclude that when we treat the system by relativistic formalism we obtain a correction factor equivalent to $K_{B}$. Finally, we expect in a near future, in agreement with experiments, that our results may be used as a useful tool to study these properties.

\section*{Acknowledgments}
\hspace{0.5cm}The authors would like to thank the Funda\c{c}\~{a}o Cearense de apoio ao Desenvolvimento Cient\'{\i}fico e Tecnol\'{o}gico (FUNCAP) the Coordena\c{c}\~ao de Aperfei\c{c}oamento de Pessoal de N\'ivel Superior (CAPES), and the Conselho Nacional de Desenvolvimento Cient\'{\i}fico e Tecnol\'{o}gico (CNPq) for financial support. R.R.S.O and A.A.A.F would like to thank Jefferson Paixão for the fruitful discussions.

\end{document}